\documentclass[conference]{IEEEtran}
\usepackage{cite}
\usepackage{graphicx,epstopdf,url}
\usepackage{subfigure}
\usepackage{bbold}
\usepackage{verbatim}
\usepackage{textcomp}
\usepackage{slashbox}
\usepackage{amsmath}
\usepackage{amssymb}
\usepackage{multirow}
\usepackage{booktabs}
\usepackage[utf8]{inputenc}  
\usepackage[lined,ruled,boxed,commentsnumbered,linesnumbered]{algorithm2e}
\usepackage{indentfirst}
\usepackage{bbm}
\usepackage[all]{xy}
\usepackage{amsmath}

\usepackage{array}
\newcolumntype{L}[1]{>{\raggedright\let\newline\\\arraybackslash\hspace{0pt}}m{#1}}
\newcolumntype{C}[1]{>{\centering\let\newline\\\arraybackslash\hspace{0pt}}m{#1}}
\newcolumntype{R}[1]{>{\raggedleft\let\newline\\\arraybackslash\hspace{0pt}}m{#1}}

\begin{document}

\title{Improving Image-recognition Edge Caches\\ with a Generative Adversarial Network}

\author{
    \IEEEauthorblockN{Guilherme B. Souza\IEEEauthorrefmark{1}, Roberto G. Pacheco\IEEEauthorrefmark{1}, Rodrigo S. Couto\IEEEauthorrefmark{1}}
   \IEEEauthorblockA{\IEEEauthorrefmark{1}Universidade Federal do Rio de Janeiro, GTA/PEE-COPPE/DEL-Poli, Rio de Janeiro, RJ, Brazil\\
    Email: bergman@gta.ufrj.br, pacheco@gta.ufrj.br, rodrigo@gta.ufrj.br}
}

\maketitle

%\IEEEpeerreviewmaketitle

\begin{abstract}

Image recognition is an essential task in several mobile applications. For instance, a smartphone can process a landmark photo to gather more information about its location. If the device does not have enough computational resources available, it offloads the processing task to a cloud infrastructure. Although this approach solves resource shortages, it introduces a communication delay. Image-recognition caches on the Internet's edge can mitigate this problem. These caches run on servers close to mobile devices and stores information about previously recognized images. If the server receives a request with a photo stored in its cache, it replies to the device, avoiding cloud offloading. The main challenge for this cache is to verify if the received image matches a stored one.
Furthermore, for outdoor photos, it is difficult to compare them if one was taken in the daytime and the other at nighttime. In that case, the cache might wrongly infer that they refer to different places, offloading the processing to the cloud. This work shows that a well-known generative adversarial network, called ToDayGAN, can solve this problem by generating daytime images using nighttime ones. We can thus use this translation to populate a cache with synthetic photos that can help image matching. We show that our solution reduces cloud offloading and, therefore, the application's latency.
\footnote{\textcopyright2022 IEEE. Personal use of this material is permitted. Permission from IEEE must obtained for all other uses, in any current or future media, including reprinting/republishing this material for advertising or promotional purposes, creating new collective works, for resale or redistribution to servers or lists, or reuse of any copyrighted component of this work in other works.} 
\end{abstract}

\section{Introduction}
\label{sec:intro}

Mobile devices commonly use cameras to extract useful information and provide users with a better understanding of their surroundings. For example, smartphone applications, such as \textit{Google Lens}\footnote{https://lens.google.com}, identify objects or landmarks targeted by the device's camera. Once the target is recognized, \textit{Google Lens} provides detailed information about it. In the case of landmarks, the application may return the location name and some information about the place (e.g., historical data). Another example is the way cameras are used in intelligent vehicles, in which decision-making processes take visual data into account~\cite{bechtel2018deeppicar, kim2016road}.
Image recognition requires high computational and energy utilization. These requirements are often incompatible with mobile devices~\cite{cuervo2010maui}. Hence, a solution is to offload image recognition to a cloud server, which does not have the same limitations. The server, in turn, processes the image and returns the result to the device.

Although sending the image to a cloud server solves resource constraint problems, it introduces communication delay. This delay increases the application's latency and can be prohibitive to real-time scenarios~\cite{bechtel2018deeppicar}. Edge computing aims to reduce this problem and other ones caused by the cloud~\cite{pacheco2021calibration}. With this paradigm, we can place computational resources on the edge of the Internet (e.g., WiFi access points and base stations). It is thus possible to reduce the communication delay and the amount of data sent to the cloud.

An edge server can deploy image-recognition (IR) caches to reduce applications' latency and network traffic. Cachier~\cite{drolia2017cachier} and FoggyCache~\cite{guo2018foggycache} both propose IR caches on the edge. Their systems intercept requests from devices before offloading to the cloud. A device can request, for example, information about a photographed location or an object recognition in an image. When an image arrives at the edge, it is compared with other images stored in the cache. The system checks if the received image is similar to any of the stored ones. If so, it returns to the user the result stored for the matched image. Reusing results in cache avoids offloading and, therefore, reduces the application's latency. If the system cannot find any similar match, the request is forwarded to the cloud. Once the cloud solves the request, the result is returned to the device and may be stored in cache for future use.

IR caches are based on the idea that different users can send a request with similar images. Various methods can assess the similarity between images. Foggycache and Cachier use feature detection algorithms such as ORB (Oriented FAST and Rotated BRIEF)~\cite{rublee2011orb} and SIFT (Scale-invariant feature Transform)~\cite{lowe1999object}. More recent approaches opt for machine-learning techniques to encode images~\cite{venugopal2018shadow}. The encodings can be later compared to obtain a similarity value between images.

Both feature detection algorithms and machine learning approaches can have poor performance when comparing images of the same place or object but under different illumination conditions~\cite{sattler2018benchmarking}. It means that an image cache may fail to reuse results from similar requests. For example, a cache that contains stored nighttime images will not perform well when receiving daytime requests. Consequently, the edge will unnecessarily offload to the cloud, even when the cache contains information about the user's request.

This paper\footnote{
Some fragments and ideas of this work are based on our preliminary paper (\url{https://sol.sbc.org.br/index.php/courb/article/view/17116}), published in Portuguese in a Brazilian workshop. This utilization is permitted by the Brazilian publisher, as seen in~\url{https://sol.sbc.org.br/index.php/indice/conduta}.} proposes a solution to make a cache system more robust to illumination variations. Specifically, we consider an urban scenario in which the cache contains nighttime images and receives daytime requests. We achieve robustness through an image translation technique, called ToDayGAN~\cite{anoosheh2019night}, which transforms real nighttime images into synthetic day images using Generative Adversarial Networks (GANs)~\cite{goodfellow2014generative}. The synthetic images are stored in the cache, increasing the odds of solving the requests on the edge server.

The experiments presented in this paper use a common dataset in the literature that contains vehicle-captured images~\cite{maddern20171}. We evaluate two different scenarios of an IR cache that receives daytime requests. One has only real nighttime images stored. The other one contains synthetic daytime images generated from these nighttime images. Results show that the storage of synthetic images reduces cloud offloading, and therefore requests' latency. Thus, our proposal can improve the latency of applications dependent on IR edge caches.

This paper is organized as follows. Section~\ref{sec:related_work} describes related work. Section~\ref{sec:imageCache} details our approach. Section~\ref{sec:methodology} describes the evaluation methodology, while Section~\ref{sec:experiments} present our results. Finally, Section~\ref{sec:conclusion} concludes the work.

\section{Related Work}
\label{sec:related_work}
An IR cache exploit requests' locality in mobile applications. This locality can be observed, for example, in image recognition applications, in which geographically close users often request content related to the same object or landmark~\cite{guo2018foggycache}. Cachier~\cite{drolia2017cachier} is an IR cache that uses ORB to detect local features from the images included in the requests. These features are then used to train a classification model. Hence, new requests can be classified as possible objects stored in the cache that reuses the stored result in the response. The reuse made possible by Cachier achieves an average latency of up to three times less than the exclusive use of the cloud. Precog~\cite{drolia2017precog} extends Cachier's proposal, spreading parts of the cache on the mobile devices and the edge. Precog can thus reduce the application's latency up to five times less than an edge-only cache. This is true since Precog can efficiently reuse the content cached on mobile devices, avoiding unnecessary requests to external edge infrastructure. 

FoggyCache~\cite{guo2018foggycache} proposes a cache system that can reuse computation for several deep neural networks (DNNs) models. To this end, FoggyCache encodes the inputs as high dimensional vectors stored in the cache. For IR applications, it uses SIFT to detect and extract features from images that are later combined to form encoding vectors. FoggyCache also develops a method to compare the cached vectors with the new inputs to identify similar requests accurately. As a result, FoggyCache reduces an application latency by a factor of 3 to 10.

Shadow Puppets~\cite{venugopal2018shadow} uses a neural network as a feature extractor instead of the traditional methods, such as ORB and SIFT. The results of Shadow Puppets show that the neural network can efficiently identify similarities between cached data and the new ones. Thus, it shows that using a neural network as a feature extractor can increase the cache's reuse. Inspired by Shadow Puppets, our work also uses a neural network-based approach for image retrieval in the cache. 

Prior works optimize the image retrieval process or cache replacement policies, reducing the required search time and cloud offloading. In contrast, our work tackles the problem of the reduced performance of an IR cache when receiving queries under very different lighting conditions. Our proposal of storing synthetic images using a ToDayGAN is thus agnostic to IR cache solutions used in the previous works.

%\textbf{Generative Adversarial Networks (GANs).} GANs have been widely employed in computer vision applications, such as image restoration~\cite{yeh2017semantic}, artificial enlargement of training sets~\cite{shrivastava2017learning} and also artificial image generation~\cite{creativeadversarialnetworks, progan}. More specifically, previous works propose GANs models to assist applications that suffer from domain adaptation problems~\cite{tzeng2017adversarial}. However, our work does not present a new GAN model. We use a GAN model to improve the performance of a caching system that has training images from one domain (i.e., nighttime images) and is tested with images belonging to another domain (i.e., daytime images).

\section{Image-recognition edge cache}
\label{sec:imageCache}

This section presents our approach by describing the cache system and detailing the procedure of generating images.

\subsection{System Overview}

\begin{figure*}[ht]
    \centering
    \includegraphics[width=0.85\linewidth]{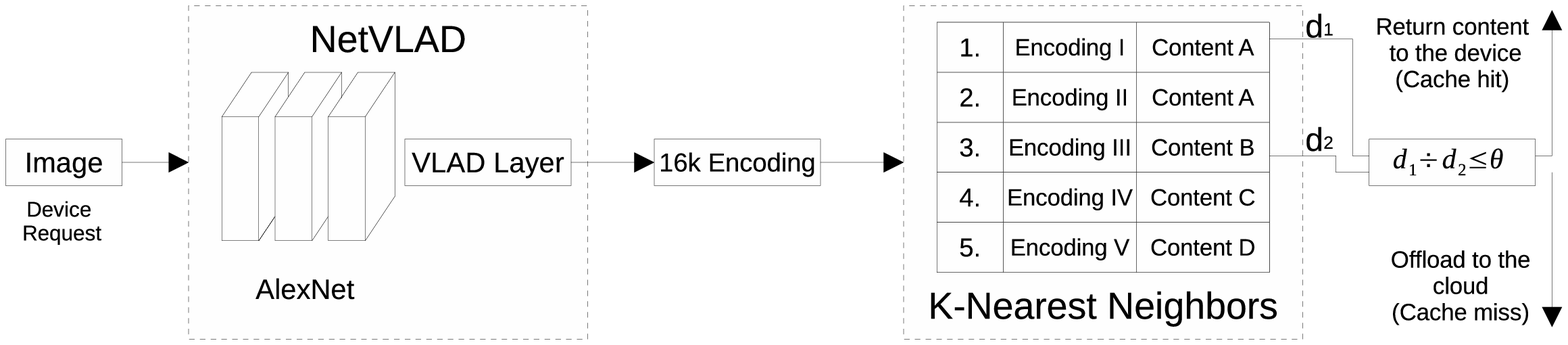}
    \caption{Our IR cache system overview. First, NetVLAD generates an encoding for the request image. This encoding is then compared with the cached ones to find the most likely requested content. The edge returns the content to the device (cache hit) or forward the request to the cloud (cache miss).}
    \label{fig:systemoverview}
\end{figure*}

An IR cache works similarly to a web cache, such as an HTTP proxy~\cite{drolia2017cachier}. A web cache stores the contents of web pages recently requested by users. Each content is indexed with an URL. The proxy looks for the request's URL to determine if the required content is stored in the cache. If so, the web cache returns the content to the user, avoiding requests to an Internet server. Otherwise, the web cache requests the desired content to the Internet server. However, IR caches do not perform an exact search with an identifier such as an URL. For example, in a geolocation application, a user sends an image of a specific location and waits for the server or cache to return information about the requested location. In this case, even if multiple users photograph the exact location, these images will have differences, such as lighting conditions or the angle at which the image is captured. The search in IR caches evaluates the similarities between requested images and the cached images extracting their descriptors using neural networks or traditional approaches, such as SIFT. These descriptors allow the cache to compute a similarity metric between the requested image and the stored ones. The reuse is then accomplished by finding similar images stored in the cache and returning the stored content.

We apply the NetVLAD~\cite{arandjelovic2016netvlad} neural network to encode the input image, extracting its descriptors. NetVLAD outputs a 16\,k dimension encoding vector for every input image. We thus compute the similarity between the encoding vector of the input image and the cached ones using Euclidean distance as the similarity metric. 
Rather than storing images themselves, the cache stores their encoding and their associated content. 

Figure~\ref{fig:systemoverview} illustrates our system. The edge receives an image sent from a mobile device and extracts its encoding using NetVLAD. This encoding is thus used as input to a k-nearest-neighbors search~\cite{JDH17}. This search finds the $k$ most similar encoding stored in the cache. Before returning the content of the most similar image, the system performs a threshold comparison to assess the correctness of the result. This comparison is based on Lowe's Ratio Test~\cite{lowe2004distinctive}. To this end, we obtain the distances $d_{1}$ and $d_{2}$, the euclidean distances between the requested image's encoding and two close ones. In Figure~\ref{fig:systemoverview}, we show an example with the five ($k=5$) closest encodings. The closest encoding is $d_{1}$. In our example, the two closest encodings have the same content $A$. Distance $d_{2}$ is hence obtained from the third closest encoding. Afterwards, we compute the ratio $r=\frac{d_{1}}{d_{2}}$ and compare it with a predefined threshold $\theta$. If $r \leq \theta$, the search result is considered incorrect, so the IR cache offloads the image to the cloud (cache miss). Otherwise, the IR cache returns the content associated with the most similar image in the cache (cache hit). We employ this system in a geolocation application. In this case, the contents are strings indicating the requested images' location.

The cloud uses a mechanism similar to the cache to reply to requests. However, the cloud has more images than the edge and may use more sophisticated image retrieval approaches. As our focus is only the IR cache, we emulate cloud image retrieval as follows. First, the cloud contains both daytime and nighttime images. Furthermore, the cloud has the corresponding encoding for the same photos used in the requests. It can be an unpractical situation but helps emulate a better image retrieval performed in the cloud. The images on the edge cache are different from the ones on the requests. Thus, the cloud can achieve much higher precision than the cache. In real scenarios, the cloud may not have access to the complete dataset beforehand and, therefore, will resort to another approach to achieve high precision. In \cite{guo2018foggycache}, for example, the cloud runs an accurate deep neural network.

\subsection{Image retrieval}

Image retrieval can use manually crafted feature extractors. These methods work by locating key points in the image and generating vectors (descriptors) to describe them. The similarity is hence determined by computing a distance between images' descriptors, such as Euclidean distance. 

Similar to Shadow Puppets, we use a neural network as an alternative to manually crafted feature extractors. Instead of extracting multiple features from an image, we use NetVLAD \cite{arandjelovic2016netvlad} to generate a single encoding descriptor for the whole image. Analogous to ORB and SIFT descriptors, the Euclidean distance between the encodings is used to determine the similarity metric between images.

NetVLAD combines both hand-engineered and machine learning approaches to propose an architecture for accurate place recognition. It consists of a convolutional neural network (CNN) connected at its end to a VLAD (Vector of Locally Aggregated Descriptors) layer, inspired by the VLAD descriptor aggregator commonly used in the image retrieval literature. They train the CNN to generate descriptors with small Euclidean distances for geographically close images and vice versa. This training allows easy comparison of NetVLAD descriptors in image retrieval. The VLAD layer of NetVLAD can be easily plugged into any CNN to form different versions of the architecture. In this work, we use the AlexNet version of NetVLAD, although a VGG-16 can achieve better results~\cite{arandjelovic2016netvlad}. We choose the AlexNet as a pessimistic approach since it can better fit hardware-constrained edge caches with low latency requirements~\cite{bianco2018benchmark}. Our training follows the same methodology of~\cite{arandjelovic2016netvlad}, using the Pittsburgh (Pitts250k)~\cite{torii2013visual} dataset.

NetVLAD has significant improvements in image retrieval in diverse databases compared to baseline models~\cite{arandjelovic2016netvlad}. Still, Anoosheh~\textit{et al.}~\cite{anoosheh2019night} and our paper show that image translation techniques improve, even more, the performance of image retrieval by storing synthetic daytime images in the database.

\subsection{Generation of synthetic images}

Generative Adversarial Networks (GANs)~\cite{goodfellow2014generative} is an unsupervised learning algorithm designed to solve generative modeling problems. In generative modeling, the training samples given by $\boldsymbol{x}$ are drawn from an unknown distribution $p_{\text{data}}(\boldsymbol{x})$. The goal of a generative modeling algorithm is to learn a probability distribution $p_{\text{model}}(\boldsymbol{x})$ that approximates $p_{\text{data}}(\boldsymbol{x})$. 
%To this end, GAN learns based on simultaneous training between two competing antagonistic DNNs. In other words, the training of GANs can be described as a game in which the two DNNs are opposing players with conflicting goals. One of the players is called generator $\mathcal{G}$, while the other is called discriminator $\mathcal{D}$. The generator model is DNN trained to generate increasingly realistic images from a random vector $\mathbf{z}$ in order to fool its adversary model, $\mathcal{D}$. In contrast, $\mathcal{D}$ learns to distinguish between real images and synthetic images generated by $\mathcal{G}$. 
GANs have recently achieved considerable advances in computer vision applications, such as image restoration~\cite{yeh2017semantic}, synthetic image generation~\cite{creativeadversarialnetworks, progan}, and image-to-image translation~\cite{isola2017image}. The latter consists of learning data distributions to convert from a source domain $X$ to a target domain $Y$. For example, a CycleGAN~\cite{zhu2017unpaired} model learns a painter's style, using a collection of their pictures, and can convert an ordinary photo to the learned style. 
%Therefore, image-to-image translation can be described as follows: we can train a GANs model to map $\mathcal{G}: X \rightarrow Y$. The output $\hat{y}=\mathcal{G}(x)$, where $x\in X$, is indistinguishable from $y \in Y$, when evaluated by the $\mathcal{D}$~\cite{isola2017image}. Moreover, previous works used image-to-image translation to assist applications that suffer from domain adaptation problems~\cite{tzeng2017adversarial}, such as the cache illumination problem presented in the Section~\ref{sec:intro}. However, our work does not present a new GANs model. We instead use a GANs model to improve the performance of a caching system that has training images from one domain (i.e., nighttime images) and is tested with images belonging to another domain (i.e., daytime images).

Our work employs a ToDayGAN~\cite{anoosheh2019night} to show that GANs can improve IR caches. ToDayGAN is an image-to-image translation model, performing the translation between daytime and nighttime image domains. This model, which has a structure based on a CycleGAN, is trained on an urban images dataset gathered by an autonomous vehicle day and night.
%Our work also uses this dataset, detailed latter in Section~\ref{sec:methodology}. 

ToDayGAN design aims to aid geolocation image retrieval~\cite{anoosheh2019night}. ToDayGAN's proposal considers that the reference database stores daytime images and that the system receives nighttime photos. Thus, before searching for the most similar image into the database, ToDayGAN transforms nighttime images into synthetic daytime images to improve the performance. Next, ToDayGAN extracts features of the transformed image to compare with the stored ones and finds the best match. Our work applies ToDayGAN in IR caches also with geolocation application in mind. We exploit the ToDayGAN's capability of matching images under different illumination conditions to reduce cloud offloading and the latency in cache-dependent applications.
Our work follows the same idea used in ToDayGAN~\cite{anoosheh2019night}, considering turning daytime images into nighttime ones. Moreover, we translate the images already stored in the IR cache instead of translating the photos in real-time. This translation should occur when the cache server is idle to not affect the system's performance. 

\section{Methodology}
\label{sec:methodology}

This work uses the Oxford RobotCar dataset~\cite{maddern20171}, composed of frames from several videos recorded on the same route in Oxford, England. An autonomous vehicle gathers these frames through cameras attached to its front, back, and sides. This vehicle repeats the same path for several days, recording different weather and lighting conditions. The dataset is divided into days of recording. We then sample images from one recording with overcast weather and use them as daytime requests. Images from two different night recordings are sampled to form the cached content. We use the ToDayGAN model trained in~\cite{anoosheh2019night} for the image-to-image translation. It also uses the Oxford RobotCar, employing part of this dataset for its training. Although our experiments use the same dataset, the images used to train this model are not part of the test set for evaluating our proposal.

Each image of Oxford RobotCar has GPS data indicating its coordinates.
Therefore, we group the dataset images by location to consider a geolocation application. The vehicle's route is split into segments of five meters each. The experiments assume that images that originate from the same segment belong to the same place and hence should receive the same result from the application when sent as a request.

Our experiments consider two main scenarios. In the first one, referred to as ``Night'', the cache stores only nighttime images. The other one, referred to as ``GAN'', evaluates the performance of the same cache system but instead stores only synthetic daytime images generated by the ToDayGAN. These synthetic images are generated through the translation of the nighttime photos used in the first scenario. In both cases, the cache only receives real daytime images as requests. 

We evaluate three metrics. The precision is the percentage of requests that the cache or the cloud have correctly identified their images' location. The recall is the percentage of correctly identified requests whose content is cached. Finally, we evaluate the application's latency, described later in Section~\ref{sec:latency}. 

\section{Results}
\label{sec:experiments}

This section evaluates our proposal. We use the experiments of Section~\ref{sec:threshold} to choose an adequate threshold $\theta$ value. Afterward, we analyze in Section~\ref{sec:cachedPlaces} the impact of the number of cached places in precision and recall. Finally, Section~\ref{sec:latency} evaluates the benefits of our proposal to a geolocation application's latency.

\subsection{Threshold choice}
\label{sec:threshold}

%This first experiment considers an IR cache that stores images from 50 different places.
This first experiment evaluates the IR cache for different thresholds, using the ``Night'' and ``GAN'' scenarios and the case where no cache is available. For each threshold, we perform ten experimental rounds. In each round, we choose 100 possible places from the dataset. We then choose to cache 50\% of these places.
This percentage allows verifying if the chosen thresholds correctly discard results of images belonging to places not cached. For each cached place, we have eight stored encodings corresponding to eight different images. We choose this value arbitrarily to show that our solution is still necessary even when the cache stores many images. In this first experiment, we use 150 images requests, chosen randomly from the 100 possible locations of each round. All results of this work are average values of ten rounds with confidence intervals of 95\%.

\begin{figure}[ht]
\center
\subfigure
[Precision.]
{\label{fig:thresholdxprecision}\includegraphics[width=0.6\linewidth]{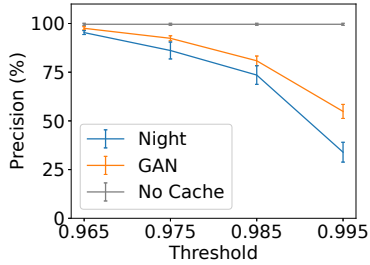}}
\subfigure
[Recall]
{\label{fig:thresholdxrecall}\includegraphics[width=0.6\linewidth]{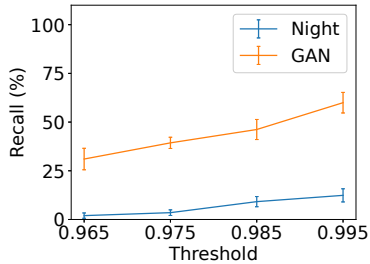}}
\caption{Threshold variation with 50\% of cached places.}
\label{fig:thresholdAdjustment}
\end{figure}

Figure~\ref{fig:thresholdxprecision} shows the precision results that indicate the percentage of requests in which the system retrieves the correct place. An incorrect image retrieval occurs when the cache returns the information of a location different from the requested one. This false retrieval also occurs when the edge offloads a request to the cloud, and the cloud returns the wrong place. This last case is rare since the cloud performs an almost ideal image retrieval, as described in Section~\ref{sec:imageCache}. To illustrate this behavior, the ``No Cache'' curve of Figure~\ref{fig:thresholdxprecision} shows the precision when the edge offloads all requests to the cloud, and it is not dependent on the threshold.  

For the ``Night'' and ``GAN'' scenarios, Figure~\ref{fig:thresholdxprecision} shows that the precision shifts down as we increase the threshold value. The recall curve of Figure~\ref{fig:thresholdxprecision} represents the percentage of requests that the cache returns the correct content, considering only those whose corresponding place is cached~\cite{drolia2017cachier}. Note that Figures~\ref{fig:thresholdxprecision}~and~\ref{fig:thresholdxprecision} have an opposite behavior regarding the threshold. It happens because a small threshold results in a small number of erroneous results being returned by the cache and hence a high precision. At the same time, it also forces unnecessary offloading of correct results to the cloud, lowering the recall. Changing this value thus allows fine-tuning the trade-off between recall and precision. 

To achieve a high recall while maintaining a high precision (around 90\%), we set the threshold value to 0.975 for the remaining experiments. It is also important to note that, although the ``Night'' scenario reaches a precision close to the ``GAN'', 
%the same values of precision in Figure~\ref{fig:thresholdxprecision}, 
this last one achieves a higher recall. It is the first evidence that the image translation technique can increase an IR cache performance by reducing the amount of offloaded requests to the cloud. The following experiments confirm the technique's efficiency for a different number of cached places.

\subsection{Cache coverage variation}
\label{sec:cachedPlaces}

The following experiment uses a similar methodology to the first one. This time, we fix the threshold value at 0.975 and vary the number of places stored in the cache. This number can be seen as cache coverage. A cache that stores images from many locations can receive requests from a larger geographical area than a cache that stores fewer places. Depending on its configuration, a cache may cover nearby streets, neighborhoods, cities, etc. As in Section~\ref{sec:threshold}, the cache has 50\% of all possible locations with eight images for each. Hence, for every number of cached places, we set the number of possible locations accordingly. As in Section~\ref{sec:threshold}, the total number of requests is three times the number of cached locations. Note that for 50 cached places, we have the same experiment of Section~\ref{sec:threshold} when $\theta=0.975$, with 100 possible locations and 150 requests.  

\begin{figure}[ht]
\center
\subfigure
[Precision.]
{\label{fig:precision}\includegraphics[width=0.49\linewidth]{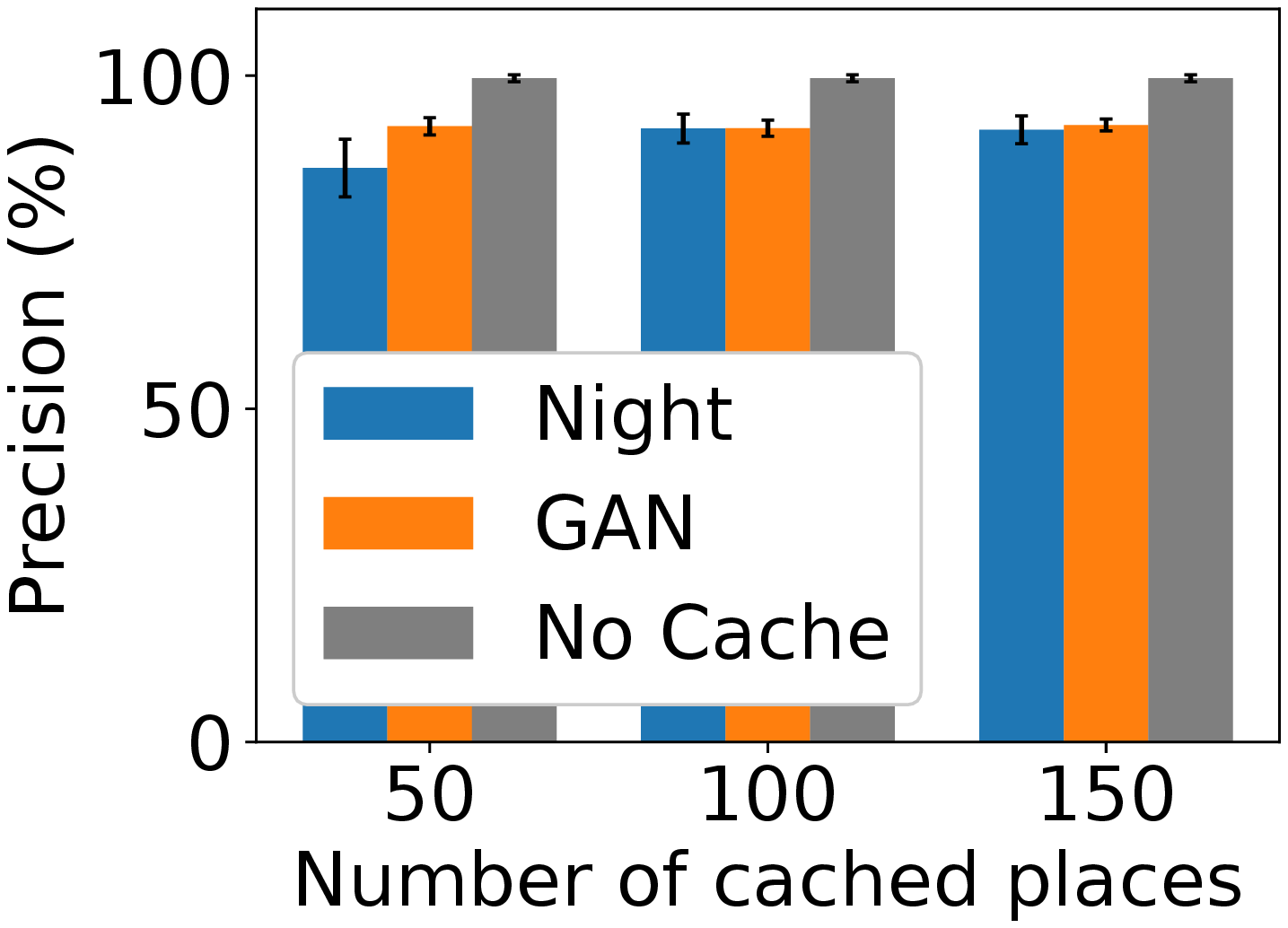}}
\subfigure
[Recall.]
{\label{fig:recall}\includegraphics[width=0.49\linewidth]{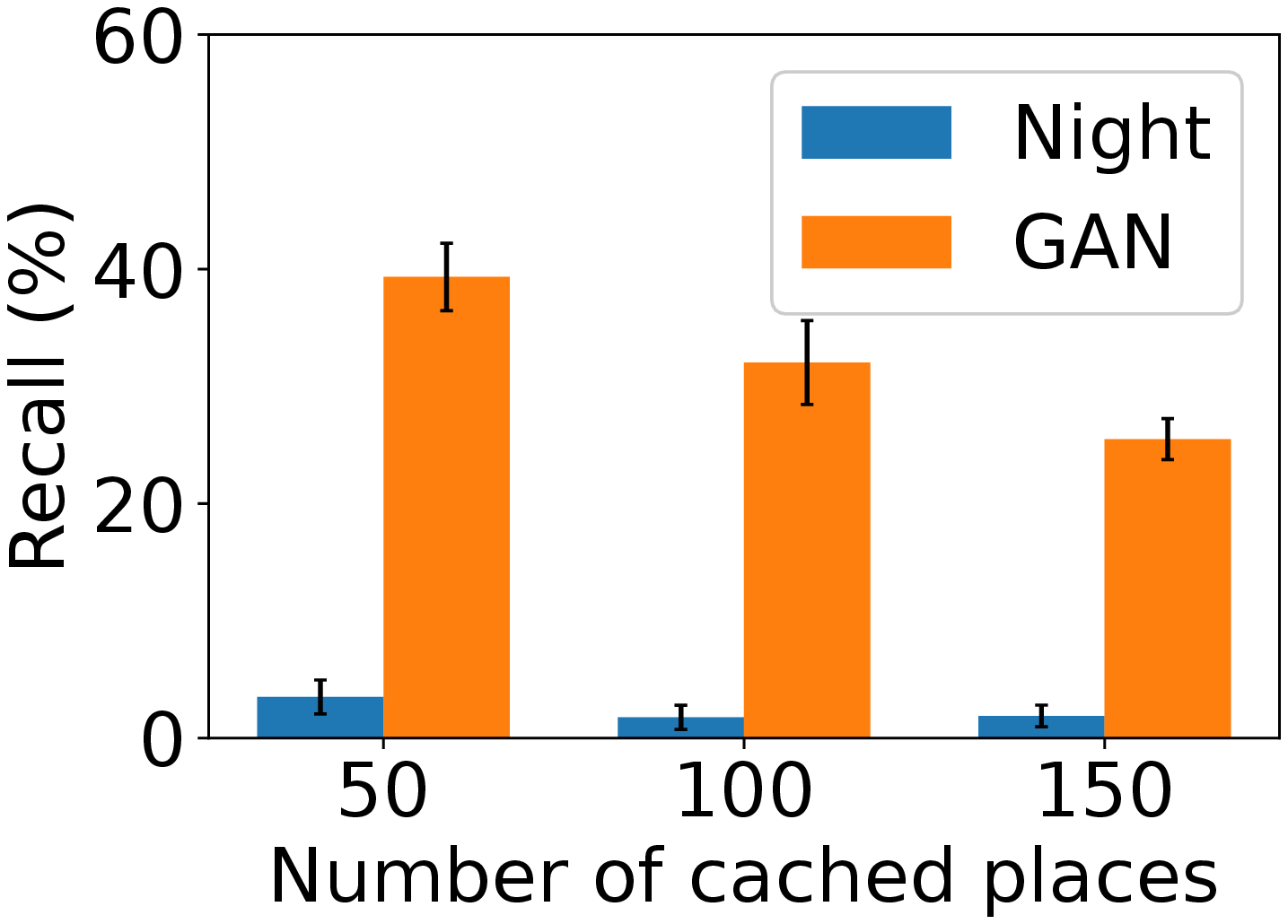}}
\caption{Cache coverage variation with 50\% of cached places and $\theta=0.975$.}
\label{fig:precisionAndRecall}
\end{figure}

Figure~\ref{fig:recall} shows that the recall decays as we increase the number of cached places. We expect this behavior because the cache needs to distinguish an image between more possibilities when more locations are cached. On the other hand, Figure~\ref{fig:precision} shows that altering the number of cached places causes no significant impact on the precision. 
Finally, as in the first experiment, the cache with synthetic images achieves a higher recall for any configuration. We fix the number of cached places in 50 for the subsequent evaluation to represent a small edge cache that receives only nearby requests.

\subsection{Latency evaluation}
\label{sec:latency}

The final experiment evaluates the impact of the recall improvements on the latency. We use the same methodology as before, but now we measure the latency. We define this metric as the total time elapsed since the arrival of a request on the edge until the obtention of its response. We disregard the communication delay between the device and the edge because it is agnostic to the cache solution. 

We set up the experiment with two machines working as the edge cache and the cloud. The cache runs a Debian 9.13 operating system. It has an Intel i5-9600K CPU with six cores at 3.70 GHz and an NVIDIA GeForce RTX 2080 Ti GPU. The cloud server is an Amazon AWS EC2 \texttt{g4dn.xlarge} instance, with a Deep Learning AMI (Ubuntu 18.04) 51.0, four vCPUs from an Intel Cascade Lake CPU, and an NVIDIA Tesla T4 GPU. The edge runs in our laboratory in Rio de Janeiro, Brazil. We instantiate the cloud machine in two different AWS regions to consider different network conditions. One region is far from the edge, located in Ohio, USA. The other one is in S\~ao Paulo, Brazil, the closest AWS region to Rio de Janeiro at the date of this paper's writing.

Table~\ref{tab:netcond} shows the network conditions between the cache and the cloud for each region. To this end, we apply \texttt{iPerf} and \texttt{ping} tools to measure the average RTT (Round-Trip Time) and throughput. This table only illustrates the network conditions. Their values depend on the Internet performance, suffering fluctuation during the day. 

\begin{table}[ht]
    \centering
    \caption{Network Conditions For Each Cloud Region.}
    \begin{tabular}{|c|c|c|c|}
    \hline
        AWS Region & Location & Throughput & RTT \\
        \hline
        us-east-2 & Ohio & 86.9\,Mbit/sec & 145.65\,ms \\
        \hline
        sa-east-1 & São Paulo & 93.4\,Mbits/sec & 12.39\,ms\\
        \hline
    \end{tabular}
    \label{tab:netcond}
\end{table}

\begin{figure}[h]
\center
\subfigure
[Ohio (USA) - us-east-2.]
{\label{fig:latencyOhio}\includegraphics[width=0.6\linewidth]{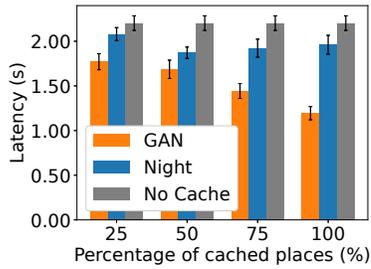}}
\subfigure
[São Paulo (Brazil) - sa-east-1]
{\label{fig:latencySP}\includegraphics[width=0.6\linewidth]{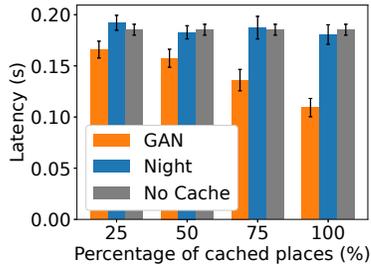}}
\caption{Latency evaluation for $\theta$ = 0.975 and 50 cached places.}
\label{fig:responseTime}
\end{figure}

Figure~\ref{fig:responseTime} presents the latency results for the two AWS regions. In the x-axis, we vary the percentage of cached places. The previous experiments always fixed this percentage in 50\%. We now set the number of cached places to 50 and change the number of possible locations to induce each percentage. This value is varied since it has a direct impact on latency. If the cache has a higher fraction of the total locations, it will result in less offloading to the cloud and, therefore, lower latency. This percentage changes over time in a production system and depends on the cache's replacement policy, which is orthogonal to our proposal. The results show that storing synthetic daytime images reduces the application's latency even for very different network conditions. This reduction is more expressive as we increase the percentage of requests belonging to cached places. Therefore fewer requests are offloaded to the cloud. Figure~\ref{fig:latencySP} shows the results for the region with the lowest offloading penalty. Even in this case, the GAN reduces the latency from 13.5\% to 40.6\% compared to using real nighttime images.

\section{Conclusions}
\label{sec:conclusion}

This paper has shown that an image retrieval (IR) cache containing nighttime images exhibits low performance when receiving daytime images as requests. We solve this problem using ToDayGAN~\cite{anoosheh2019night}, a well-known GAN-based image generator, to translate the IR cache's content into daytime images. With this approach, the cache can reuse more computation, avoiding cloud offloading. Our experiments have used an actual cloud provider to show that the proposal reduces the application's latency. This effect is thus our main benefit, being particularly useful to latency-sensitive applications as they can experience a drop in user experience. These applications may even be rendered unfeasible by high response times. For future works, we plan on evaluating our proposal considering other concerns typical of cache literature, such as replacement policy. Furthermore, our solution is independent of the cache implementation. Hence, we may also investigate the image translation technique efficiency for different datasets and image retrieval mechanisms.

\section*{Acknowledgements}
This study was financed in part by the Coordenação de Aperfeiçoamento de Pessoal de Nível Superior - Brasil (CAPES) - Finance Code 001. It was also supported by CNPq, FAPERJ Grants E-26/203.211/2017, E-26/211.144/2019, and E-26/201.300/2021, and FAPESP Grant 15/24494-8.
\bibliographystyle{IEEEtran}
\bibliography{header,bibFile}

\end{document}